\documentclass[journal]{IEEEtran}
\usepackage[numbers,sort&compress]{natbib}

\usepackage[cmex10]{amsmath}
\usepackage{array}
 \usepackage[caption=false,font=normalsize,labelfont=sf,textfont=sf]{subfig}
\usepackage{float}
\usepackage{graphicx}

\begin{document}
%
% paper title
% can use linebreaks \\ within to get better formatting as desired
% Do not put math or special symbols in the title.
\title{A Global Analysis of Light and Charge Yields in Liquid Xenon}
%
%
% author names and IEEE memberships
% note positions of commas and nonbreaking spaces ( ~ ) LaTeX will not break
% a structure at a ~ so this keeps an author's name from being broken across
% two lines.
% use \thanks{} to gain access to the first footnote area
% a separate \thanks must be used for each paragraph as LaTeX2e's \thanks
% was not built to handle multiple paragraphs
%

\author{Brian~Lenardo,~\IEEEmembership{Member,~IEEE},
        Kareem~Kazkaz,
        Aaron~Manalaysay,
        Jeremy~Mock,
        Matthew~Szydagis,
        and~Mani~Tripathi %,~\IEEEmembership{Life~Fellow,~IEEE}% <-this % stops a space
\thanks{B. Lenardo, A. Manalaysay, and M. Tripathi are with the University of California - Davis, Davis,
CA, 95616 USA (e-mail: bglenardo@ucdavis.edu, aaronm@ucdavis.edu, mani@physics.ucdavis.edu).}% <-this % stops a space
\thanks{B. Lenardo and K. Kazkaz are with Lawrence Livermore National Laboratory, Livermore, CA 94550 USA (email: kazkaz1@llnl.gov).}% <-this % stops a space
\thanks{J. Mock and M. Szydagis are with the Department of Physics University at Albany - SUNY, Albany, NY, 12222 USA (email: jmock@albany.edu, mszydagis@albany.edu).}}
%\thanks{Manuscript received April 19, 2005; revised December 27, 2012.}

% note the % following the last \IEEEmembership and also \thanks - 
% these prevent an unwanted space from occurring between the last author name
% and the end of the author line. i.e., if you had this:
% 
% \author{....lastname \thanks{...} \thanks{...} }
%                     ^------------^------------^----Do not want these spaces!
%
% a space would be appended to the last name and could cause every name on that
% line to be shifted left slightly. This is one of those "LaTeX things". For
% instance, "\textbf{A} \textbf{B}" will typeset as "A B" not "AB". To get
% "AB" then you have to do: "\textbf{A}\textbf{B}"
% \thanks is no different in this regard, so shield the last } of each \thanks
% that ends a line with a % and do not let a space in before the next \thanks.
% Spaces after \IEEEmembership other than the last one are OK (and needed) as
% you are supposed to have spaces between the names. For what it is worth,
% this is a minor point as most people would not even notice if the said evil
% space somehow managed to creep in.

% The paper headers
\markboth{}%
{Shell \MakeLowercase{\textit{et al.}}: Bare Demo of IEEEtran.cls for Journals}
% The only time the second header will appear is for the odd numbered pages
% after the title page when using the twoside option.
% 
% *** Note that you probably will NOT want to include the author's ***
% *** name in the headers of peer review papers.                   ***
% You can use \ifCLASSOPTIONpeerreview for conditional compilation here if
% you desire.

% If you want to put a publisher's ID mark on the page you can do it like
% this:
%\IEEEpubid{0000--0000/00\$00.00~\copyright~2012 IEEE}
% Remember, if you use this you must call \IEEEpubidadjcol in the second
% column for its text to clear the IEEEpubid mark.

% use for special paper notices
%\IEEEspecialpapernotice{(Invited Paper)}

% make the title area
\maketitle

% As a general rule, do not put math, special symbols or citations
% in the abstract or keywords.
\begin{abstract}
We present an updated model of light and charge yields from nuclear
recoils in liquid xenon with a simultaneously constrained parameter set.
A global analysis is performed using measurements of electron and photon
yields compiled from all available historical data, as well as measurements of the
ratio of the two. These data sweep over energies from 1 - 300 keV and external applied 
electric fields from 0 - 4060 V/cm. The model is constrained by constructing global cost 
functions and using a simulated 
annealing algorithm and a Markov Chain Monte Carlo approach to optimize and find confidence 
intervals on all free parameters in the model. This analysis contrasts with previous work in that we
do not unnecessarily exclude data sets nor impose artificially conservative assumptions, do not use spline functions,
and reduce the number of
parameters used in NEST v0.98. We report our results and the calculated best-fit charge 
and light yields. These quantities are crucial to understanding the response of 
liquid xenon detectors in the energy regime important for 
rare event searches such as the direct detection of dark matter particles.
\end{abstract}

% Note that keywords are not normally used for peerreview papers.

% For peer review papers, you can put extra information on the cover
% page as needed:
% \ifCLASSOPTIONpeerreview
% \begin{center} \bfseries EDICS Category: 3-BBND \end{center}
% \fi
%
% For peerreview papers, this IEEEtran command inserts a page break and
% creates the second title. It will be ignored for other modes.
\IEEEpeerreviewmaketitle

% The very first letter is a 2 line initial drop letter followed
% by the rest of the first word in caps.
% 
% form to use if the first word consists of a single letter:
% \IEEEPARstart{A}{demo} file is ....
% 
% form to use if you need the single drop letter followed by
% normal text (unknown if ever used by IEEE):
% \IEEEPARstart{A}{}demo file is ....
% 
% Some journals put the first two words in caps:
% \IEEEPARstart{T}{his demo} file is ....
% 
% Here we have the typical use of a "T" for an initial drop letter
% and "HIS" in caps to complete the first word.

\section{\label{sec:Introduction}Introduction}

\IEEEPARstart{L}{iquid} xenon is currently of great interest in the detection
and measurement of ionizing radiation. Applications under study include research in direct dark matter detection,
neutrino physics, nuclear non-proliferation, and medical imaging \cite{LXeReview, LXeReview2, Knoll, AprileBook}. Due to the wide
application of the technique, it is important to develop a detector-independent understanding of how the medium
responds to incident radiation.

The Noble Element Simulation Technique (NEST) incorporates a semi-empirical physical model of the generation of 
scintillation photons and ionization electrons from recoiling particles in liquid xenon \cite{NESTpaper, NESTpaper2, JeremyPaper}.  In both argon \cite{MikeFoxe} and helium \cite{DanMHelium}, it is possible to calculate excitation and ionization in recoil cascades from first principles using measurements of the relevant interaction cross-sections.  In xenon these cross sections have never been measured or calculated, rendering such predictions impossible. NEST is intended to provide a standardized way to predict yields in the absence of such information. Moreover, the underlying model is continuously compared to measurements to ensure agreement with experiment.  The NEST software is built for easy integration into the Geant4 package \cite{Geant4_1, Geant4_2}, allowing the simulation and prediction of detector responses using standard Monte Carlo techniques. While other software exists to model ionization, scintillation, or recoil tracks, there is no comprehensive package that models both ionization and scintillation as a function of both energy and applied electric field.  NEST can be used by the larger community to compare to new measurements and interpret experimental results \cite{MIXPaper, QingLin, XENON100Result, LUXresult}.

Of particular interest to particle physics applications is the ability to discriminate between electronic recoils
(ER) resulting from $\gamma$ and $\beta$ radiation and nuclear recoils (NR) produced by massive neutral particles. In both  dark matter searches and searches for coherent neutrino-nucleus scattering, low energy nuclear recoils constitute the expected signal while low energy electronic recoils constitute backgrounds. 
Discrimination between the two is often accomplished in dual-phase time projection chambers (TPCs) by measuring both the scintillation signal
produced by excited xenon molecules and the charge signal produced by ionization of the xenon atoms 
\cite{LUXresult,XENON100Result}. The ratio of these two signals differs between ER and NR events, allowing particle-type discrimination.  Thus it is important to be able to accurately predict scintillation and ionization yields not only for energy reconstruction, but to understand background rejection in such experiments as well.

In this work, we improve the modeling of nuclear recoils
at energies below 300 keV and develop a new method to validate models in the context of a large body of calibration measurements. 
We begin by explaining the physical interpretation and parameterization
of our model, then constrain the model using a plethora of published experimental data.
We end with a discussion of our results and their application in understanding the yields of NR 
interactions in the liquid.

\begin{figure}[h]
\centering
\includegraphics[width=0.49\textwidth]{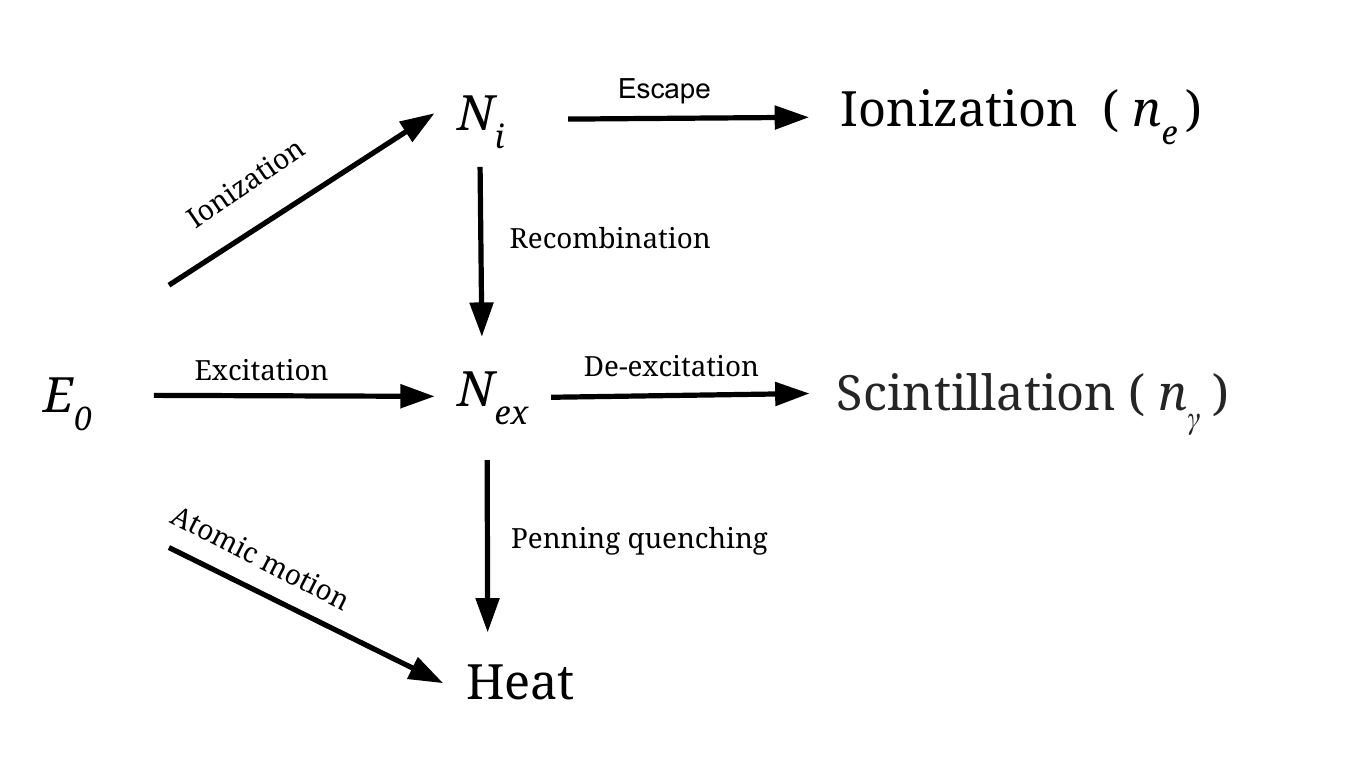}
\caption{A schematic of the process by which an energy deposition in liquid xenon produces photons and electrons ($n_{ph}$ and $n_e$)}
\label{fig:EnergyPartition}
\end{figure}

\section{\label{sec:Model} The Nuclear Recoil Model}
\subsection{\label{subsec:Theory}Theoretical Framework}
The observable quantities produced by an energy deposition in a liquid xenon detector are the scintillation photons ($n_{ph}$) and the ionization electrons ($n_e$).  The transmutation of deposited energy into these quanta is governed by the processes described in this section.  

The model used in NEST is constructed from a simple physical picture of the
process of a particle depositing energy in liquid xenon, sketched in Fig. \ref{fig:EnergyPartition}. An energy deposition of $E_0$ in the medium
is distributed between two measurable channels: formation of excitons ($N_{ex}$) and formation of electron-ion pairs ($N_i$). Some energy 
is additionally lost as unmeasurable dissipation of heat. The process determines the number of quanta, $n_q$, produced by an energy deposition from the 
conservation of energy according to a simplified version of Platzman's equation for rare gases \cite{Platzman}. For nuclear recoils, a quenching
factor $L$ is applied to account for the energy lost to atomic motion rather than the
detectable electronic channels:
\begin{equation}
\begin{aligned}
 n_q &= \frac{E_0L}{W}\\
 n_q &= N_{ex} + N_i
 \end{aligned}
\end{equation}
In the above, $W$ is the empirically-determined average energy required to produce a quantum (either an exciton or ion) in the liquid.
This quantity includes the energy lost to sub-excitation electrons, and may be higher than the
actual energy required to produce quanta.

The quenching factor $L$ is given by Lindhard's
theory as described in \cite{Lindhard} and \cite{SorensenDahl}, with 
\begin{equation}
 L = \frac{k \, g(\epsilon)}{1 + k\, g(\epsilon)}
\end{equation}
where $k$ is a proportionality constant between the electronic stopping power and the velocity of the recoiling
nucleus. The quantity $g(\epsilon)$ is proportional to the ratio of electronic stopping power to nuclear stopping power.
It is a function of the energy deposited, usually converted to the dimensionless quantity $\epsilon$ with
\begin{equation}
\epsilon = 11.5 (E_0 / keV) Z^{-7/3}
\end{equation}
where $Z$ is the atomic number of the nucleus.
In these terms, $g(\epsilon)$ is given in \cite{LewinSmith} by
\begin{equation}
g(\epsilon) = 3 \epsilon^{0.15} + 0.7 \epsilon^{0.6} + \epsilon
\end{equation}
While equations 2 and 3 are valid for any nuclei, in the case of xenon ($Z = 54$), the quenching factor $L$ is important for interactions with $E_0<\,\sim10$ MeV. There exist alternative models of this quenching, and we explore these in Section \ref{sec:Discussion}.

From Eq. 1 we see that $n_q$
produced for a given $E_0$ is governed by the ratio $L/W$, so any systematic shift in $W$ can be 
offset by a corresponding shift in $L$ in the fit. We treat 
the exciton-to-ion ratio 
$N_{ex} / N_i$ as a field- and energy-dependent parameter (described in Sec.
\ref{sec:Constrain}), enabling the above 
equation to be solved to obtain the number of excitons and ions created by a given energy deposition.

While excitons lead directly to the emission of scintillation photons, the electron-ion pairs undergo 
further division via electron-ion recombination. The probability of recombination $r$ is calculated using 
the Thomas-Imel box model \cite{ThomasImel}, which gives 
\begin{equation} 
r = 1 - \frac{\text{ln}(1 + N_i \, \varsigma)}{N_i \, \varsigma}
\end{equation}
The quantity $\varsigma$ is parameterized with a power law dependence on applied electric field and fit to data. 
Free electrons that recombine with ions add to the scintillation light signal, while electrons that escape become 
the collected charge signal. 

A final quenching is applied to the light signal to account for Penning effects, in which 
two excitons can interact to produce a single photon \cite{MeiHime}. This quenching is parametrized by
\begin{equation}
f_l = \frac{1}{1 + \eta\, \epsilon\,^{\lambda} } 
\end{equation}
where $\eta$ and $\lambda$ are left as free parameters.  This expression
is given by Birk's Saturation Law, with $\eta\, \epsilon\,^{\lambda}$ proportional to the electronic stopping power. 
The result of this is an increased quenching effect with increasing energy, due to higher ionization density along
the track of the recoiling Xe atom.

The final expressions for number of electrons $n_e$ and number of photons $n_{ph}$ produced by an energy
deposition $E_0$ are
\begin{equation}
n_e = L(E_0) \times \frac{E_0}{W} 
\left(\frac{1}{1 + N_{ex}/N_i}\right) (1 - r)
\end{equation}
\begin{equation}
n_{ph} = L(E_0) \times f_l \times\frac{E_0}{W}
\left[ 1 -  \left(\frac{1}{1 + N_{ex} / N_i}\right) (1-r) \right] 
\end{equation}

\subsection{\label{subsec:Parameterizing}Parameterizing The Model}

To fit the model to data, the quantities $N_{ex}/N_i$ and $\varsigma$ are parameterized to account for dependence
of yields on applied electric field and energy. Each is treated as a power law function of an
externally applied electric field $F$, and the 
exciton-to-ion ratio is given an exponential dependence on energy:
\begin{equation} 
N_{ex}/N_i = \alpha \, F\,^{ - \zeta} \,(1 - e^{ - \beta \epsilon} ) 
\end{equation}
\begin{equation} 
\varsigma = \gamma F\, ^{ - \delta}
\end{equation}

The parameters $\alpha$, $\zeta$, $\beta$, $\gamma$, and $\delta$ are free parameters in the fit.  The functional forms of the above parameterizations are selected based on trends observed in the data
combined with qualitative physical arguments outlined below. 

The coefficient and energy-dependence for $N_{ex}/N_i$ represent the magnitude and energy dependence of the ratio of the unknown excitation and ionization cross-sections in $\text{Xe}-\text{Xe}$ collisions in the cascade.  A variety of parameterizations were studied for energy
dependence, and we find that the data are best described by an asymptotic function that decays with decreasing energy. 
The power law field-dependence of $N_{ex}/N_i$ is
chosen based on an observed trend in best-fits at different applied fields, and can be interpreted
as electric field dependence in geminate recombination (immediate recombination of electrons with their parent ions), 
which occurs on a much
shorter timescale than Thomas-Imel recombination.  

The power-law
parameterization of $\varsigma$
follows from Dahl's arguments in \cite{DahlThesis}, from which we expect a field-dependence
$\delta$ of $\text{\bf O}\,(\,-0.1\,)$. Thomas-Imel theory explicitly predicts an $F^{-1}$ dependence:
\begin{equation}
\varsigma = \frac{\alpha'}{4 a^2 u_- F}
\end{equation}
Here $\alpha'$ is a constant recombination coefficient, $a$ is the size of a box containing the track of the recoiling particle, $u_-$ is the electron mobility, and $F$ is the applied electric field. However, the geometry of the track can be altered by the presence of external fields. As in \cite{DahlThesis}, we allow the exponent $\delta$ to float to make our model more generic.

The downward trend of both field dependencies represents the increased probability of electrons being 
extracted away from their parent ions with increasing applied electric field.

Three additional parameters are allowed to float in the fit: $\eta$ and $\lambda$ from Eq. 6 and the Lindhard
$k$ in Eq. 2.  The exponent $\lambda$ is generally taken to be 0.5, as the stopping power is expected to be proportional to the velocity at low energies \cite{Bezrukov}. The coefficient $\eta$ is equivalent to Birk's constant $k_B$ multiplied by the coefficient of the stopping power \cite{MeiHime} and a theoretical value is calculated in \cite{Bezrukov} as $\eta = 3.55$, assuming $k_B = 2.015\times10^{-3} \text{g / MeV cm}^2$.  Lindhard theory predicts $k = 0.166$, but has a large uncertainty and is expected to lie anywhere between 0.1 and 0.2 \cite{Lindhard}. An updated calculation by Hitachi finds $k = 0.110$ \cite{Hitachi}. Due to the degeneracy between $k$ and $W$ and the fact that the uncertainty in $k$ is much larger than that on $W$, we fix $W$ at its measured value of $13.7$ eV in our analysis and fit $k$ to the data.

Finally, we define a nuisance parameter, $F_0$. This is a small non-zero number that can be 
used as an $F$ value in Eqs. 7 and 8 to calculate yields with no applied electric fields. At other fields, $F$ is simply the measured value.
The implementation of $F_0$ allows the model to be completely continuous in field,
improving on previous work that treated zero-field as a special case to be separately fit to data. Physically it 
can be argued to represent small fluctuations in the local field when no external drift field is applied to the liquid. Since only the magnitude and not the direction of the applied field affects the production of quanta, $F_0$ is allowed to float with the constraint that it must be greater than zero. Evidence in \cite{DahlThesis} suggests that the light yield at drift fields of order 10 V/cm are experimentally indistinguishable to the zero-field case, providing a range of allowable best fit values for this parameter.

The recent results from the SCENE collaboration suggest that a similar approach may be useful for modeling argon \cite{SCENE}.

\section{\label{sec:Constrain}Constraining The Model}
The model is constrained using three categories of data sets, each consisting of multiple measurements of
yields in liquid xenon. The absolute NR charge yield, $\mathcal{Q}_y$, is constrained using twelve
 measurements across different energies at a range of electric fields \cite{ DahlThesis, ColumbiaQy2006, 
SorensenQy2009, SorensenQy2010, XENON10Qy, ManzurQy2010, HornQy2011, Xenon100Qy2013}. The photon yield is constrained 
with seven additional measurements
\cite{ ChepelLeff1999,ArneodoLeff2000,AkimovLeff2002,AprileLeff2005,AprileLeff2009,PlanteLeff2011,ManzurQy2010}
of the parameter $\mathcal{L}_{eff}$, defined as the zero-field scintillation yield for nuclear recoils 
relative to that of the 122 keV $\gamma$-ray from $^{57}$Co.
Finally, we constrain the ratio of charge to light, $n_{e}/n_{ph}$ using measurements 
at several fields from \cite{DahlThesis}.  In total, there are 329 data points.
In the singular case of \cite{ChepelLeff1999}, uncertainty regarding the threshold motivates the exclusion of the two lowest-energy
points \cite{ChepelPoints}, for which no overlap in uncertainty exists between the lowest energy measurements and the remainder of the world's data.
Thus 327 data points are used to constrain the model.

Each data point is treated with equal weight when constraining the model, thus measurements that quote smaller uncertainties provide stronger constraints. Consequently, if uncertainties in a particular measurement are underestimated, our analysis will be biased in favor of those data. Systematic differences in energy reconstruction and resolution are not considered in this work, but would also affect how strongly a particular measurement constrains the model. 

In the case of measurements of the charge yield, electrons are typically extracted from the liquid by applying an additional electric field across the interface. This work follows the implicit assumption in each of the references that the electron
extraction efficiency in each measurement was 100\%, unless explicitly stated.  The extraction field in each experiment can be compared to the saturation point measured 
by Gushchin et al. \cite{Gushchin, Gushchin2}.    Lower extraction efficiencies would result in reported values being underestimates
for $\mathcal{Q}_y$, and this systematic normalization uncertainty on all measurements has not been considered in this work.
 
Other sources of systematic uncertainties in the measurements are inconsistently reported, making a full treatment of all sources of error 
 prohibitively difficult. The global analysis is intended to effectively average over systematic differences in the separate measurements to discern true absolute yields. 

To fit models to the global dataset, we use Monte Carlo optimization techniques. A simulated annealing algorithm is used to study the effects of different models, and a Metropolis-Hastings algorithm is used to extract the best fits and uncertainties from the preferred model. The implementation is described in detail below. Our results for the best-fit model were cross-checked using a coarse raster scan across the parameter space combined with a gradient descent minimizer to ensure that the Monte Carlo algorithms had converged on the optimum.

 \subsection{\label{subsec:SimulatedAnnealing}Simulated Annealing}
 To facilitate experimentation with different models and constraints, we perform a 
maximum likelihood optimization using simulated annealing \cite{Kirkpatrick}. This technique allows the program to make 
random jumps in the parameter space drawn from a given proposal distribution. Steps leading to regions of higher likelihood are accepted automatically, while steps leading to regions of lower likelihood are accepted with a probability that decreases with step number $t$. The acceptance probability used in this fit was
\begin{equation} 
P(\text{accept}) = \exp{ \left( \frac{\log{ \mathcal{L}_{proposed} } - \log{ \mathcal{L}_{current} }}{ T(t)} \right) }
\end{equation} 
where $\mathcal{L}$ is the likelihood and $T(t)$ is chosen to be a linearly decreasing function of the step number for simplicity. 
This technique allows the fitter to initially explore a wide range of the parameter
space and escape local extrema, settling later in a region very near the global maximum likelihood. 

In our implementation, a uniform random number $R$ in the interval [0,1] is drawn at each step, and steps leading to lower likelihoods are accepted if $R < P(\text{accept})$. To find an optimal form of $T(t)$, the slope was varied until a value was found where O(10\%) changes in the slope did not affect the outcome of the fit. 

The likelihood function used in this algorithm assumed each data point behaved according to a split 
gaussian distribution, to take into account asymmetric errors in the data set:

\begin{equation} 
\mathcal{L} (\vec{\theta} | x) = \prod\limits_i \frac{\sqrt{2}}{\sqrt{\pi} (\sigma_+ + \sigma_-)} 
\exp{ \left( \frac{ -(x_i - x_{model})^2}{2 \sigma_{+/-}^2} \right)}
\end{equation}

In the above, $\sigma_{+/-}$ are the upper and lower uncertainties on the measurements, $x_i$ is the
measured quantity, and $x_{model}$ is that calculated from the given model using the set of 
parameters $\vec{\theta}$.  

While this method is not guaranteed to locate the absolute optimal value, tuning of
$T(t)$ and the proposal distribution allows the fit to get very close. The simulated annealing 
fitter enjoys a factor of $\sim$20 improvement in computation speed over a manual scan across the 9-dimensional parameter space,
 allowing this algorithm to 
serve as a test bed for alternative models to be incorporated into NEST, as well as rapid incorporation of new data into the fit.

\subsection{\label{subsec:MCMC}Metropolis-Hastings Markov Chain Monte Carlo (MCMC)}
A Metropolis-Hastings MCMC technique was used to extend the capability of the simulated annealing fitter to extract the final means,
variances and covariances on the free parameters in our model. The Metropolis-Hastings algorithm 
\cite{Hastings} works in a manner similar to simulated annealing, but the probability of accepting 
a step with lower likelihood is given by the ratio
\begin{equation} 
P(\text{accept}) = \frac{ \mathcal{L} (\vec{\theta} | x)_{\text{proposed}} }{ \mathcal{L} 
(\vec{\theta} | x)_{\text{current}} }
\end{equation}
The values of all the parameters are recorded at each step. As the random walk progresses, it builds a sample
 of $\mathcal{L} (\theta | x) $, revealing the structure of the underlying posterior probability distribution.
By histogramming the values of any given subset of parameters, one marginalizes over the remaining
variables, enabling quick numerical estimates of variances and covariances. For a large number
of steps, these estimates can be made arbitrarily close to the true values.

\begin{table}[t]
\renewcommand{\arraystretch}{1.3}
\caption{Best fits and 68\% confidence intervals of free parameters.}
\label{tab:fits}
\centering
\begin{tabular}{|c|c|c|}
\hline
Parameter & Best Fit & 68\% conf.\\
\hline
\hline
$\alpha$ & 1.240 & +0.079 \\
  & & -0.073 \\
\hline
$\zeta$ & 0.0472 & +0.0088 \\
 & & -0.0073 \\
\hline
$\beta$ & 239 & +28 \\
 & & -8.8 \\
\hline
$\gamma$ & 0.01385 & +0.00058 \\
 & & -0.00073 \\
\hline
$\delta$ & 0.0620 & +0.0056 \\
 & &-0.0064 \\
\hline
$k$ & 0.1394 & +0.0032 \\
 & &-0.0026 \\
\hline
$\eta$ & 3.3 & +5.3 \\
 & & -0.7 \\
\hline
$\lambda$ & 1.14 & +0.45 \\
 & & -0.09 \\
\hline
$\mathcal{C}$ & 0.00555 &+0.00025 \\
& & -0.00025 \\
\hline
\end{tabular}
\end{table}

Steps are drawn from a multivariate normal proposal distribution,
which has the form
\begin{equation}
P(\vec{\theta}) = \frac{1}{ (2 \pi)^{n/2} |\mathbf{\Sigma}|^{1/2} } \exp{ 
\left( -\frac{ (\vec{\theta} - \vec{\theta}_0 )^T \mathbf{\Sigma}^{-1} 
(\vec{\theta} - \vec{\theta}_0 ) } {2} \right)  } 
\end{equation}
where $\mathbf{\Sigma}$ is the covariance matrix, $\vec{\theta}$ and $\vec{\theta}_0$ are vectors containing
proposed and current values of the free parameters, respectively, and $n$ is the number of dimensions. The algorithm's behavior is assessed by calculating the autocorrelation of each parameter at different step separations $h$, given by the formula
\begin{equation}
\rho(h) = \frac{\sum_t [(x_t - \bar{x})(x_{t+h} - \bar{x})]}{\sqrt{ \sum_t (x_t - \bar{x})^2 \sum_t (x_{t+h} - \bar{x})^2}}
\end{equation}
where $x_t$ is the value of the parameter at step $t$ and $x_{t + h}$ is the value at step $t+h$.
The resultant $\rho(h)$ is fit to an exponential function $\rho'(h) = A \text{ exp} ( - h / \tau)$ to extract the autocorrelation length $\tau$ as in \cite{RichOttThesis}.  A larger $\tau$ is a symptom of more strongly correlated samples, and to ensure unbiased sampling, one must have a $\tau$ much less than the total number of samples being analyzed.

To meet this requirement, the
proposal distribution is tuned using an iterative procedure.  
Initially, we assume that $\mathbf{\Sigma}$ has no off-diagonal elements, and tune the values by
hand to reduce the autocorrelation length of each parameter. We then set $\mathbf{\Sigma}$ equal to the covariance matrix of the resulting samples
and feed this back into the MCMC to produce a new set of samples for analysis. After two iterations, 
we obtain autocorrelation lengths of less than 1000 steps in all nine 
parameters. Our final analysis uses a data set of $3 \times 10^6$ samples, which we are confident provides a fair sample of $\mathcal{L} (\vec{\theta} | x)$.

%Similar results can be derived by assuming that ions recombine with electrons freed from other atoms with a small probability of
%re-encounter \cite{Mozumder}.

\section{\label{sec:Results}Results}
\begin{figure}[h]
\centering
\includegraphics[width=0.44\textwidth]{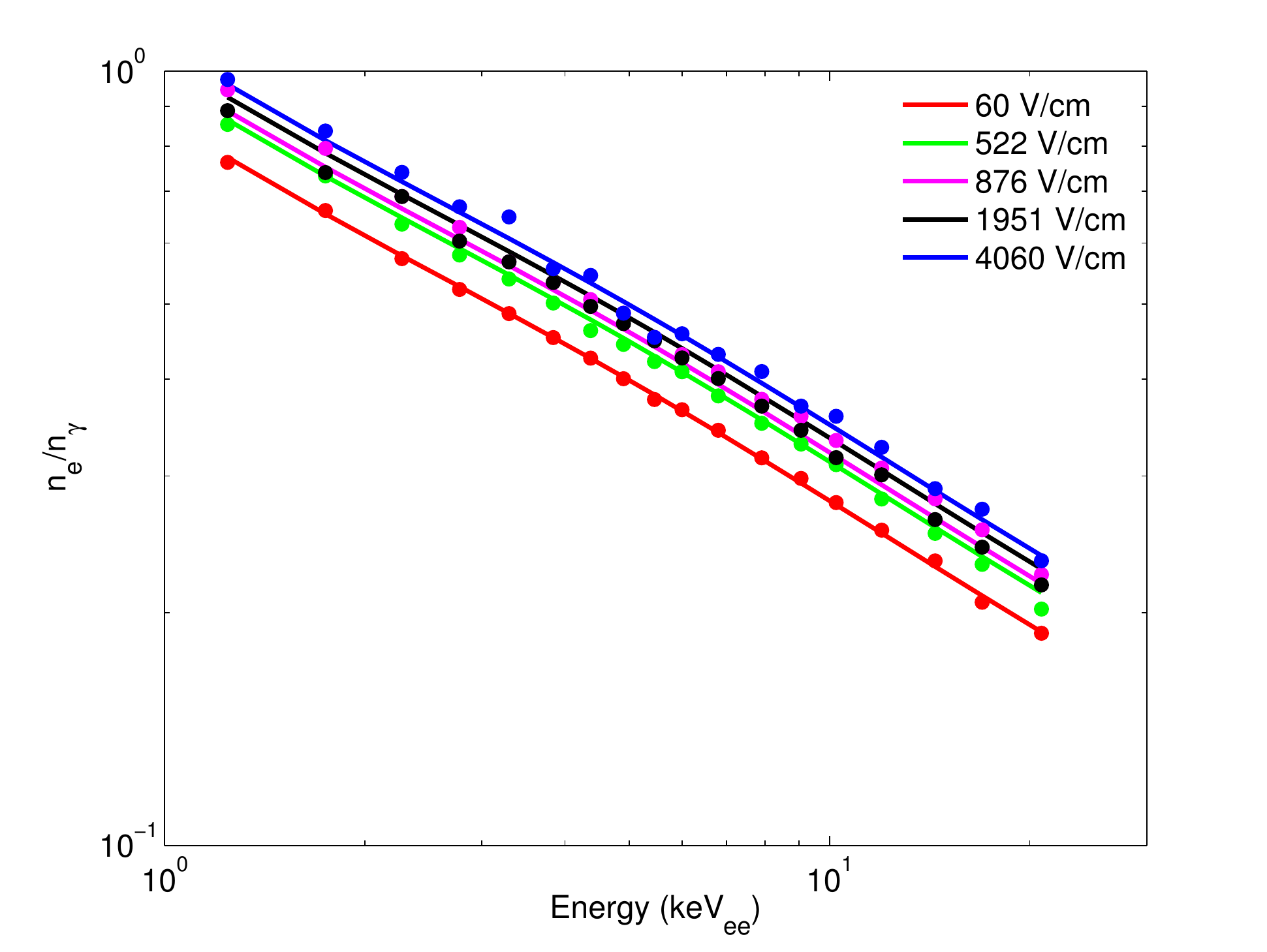}
\caption{(color online) Ratio of charge yield to light yield, compared to the data from \cite{DahlThesis} used in the 
global fit. These data constrain the exciton to ion ratio $N_{ex}/N_i$ as well as the fraction of ionization electrons that recombine with their parent ions $r$. In this plot and in Fig \ref{fig:Fluct}, we use the energy scale keV$_{ee}$, which is the electron-equivalent energy scale.
For electronic recoils, a negligible amount of energy is lost to atomic motion, corresponding to $L=1$ in Eq. 1.  Error bars are not shown because they are smaller than the data points, but are included in the fits.}
\label{fig:DahlPlot}
\end{figure}

\begin{figure}[t]
\centering
\includegraphics[width=0.49\textwidth]{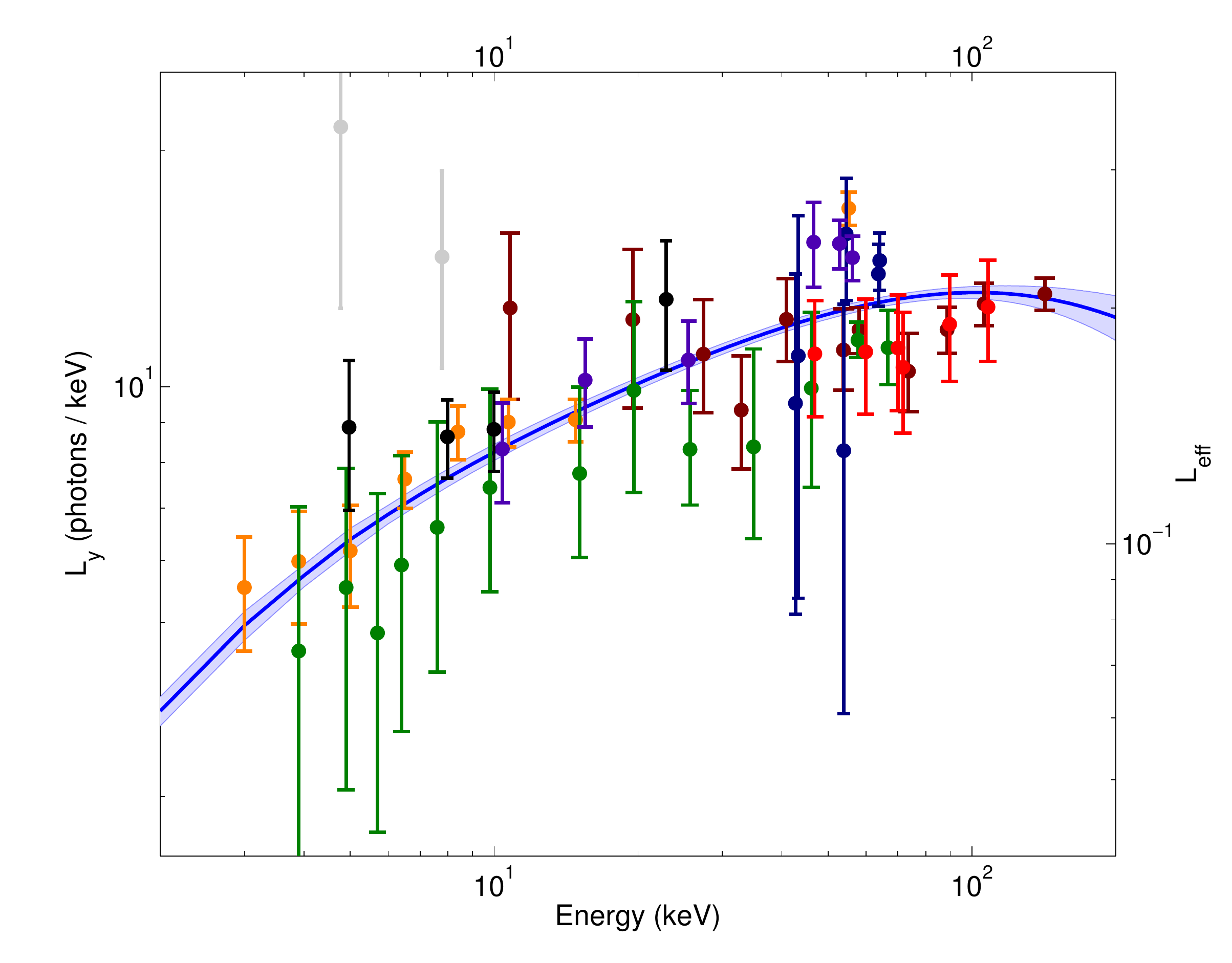}
\caption{(color online) Best-fit to $\mathcal{L}_{eff}$ (solid blue line) with statistical error band. Also shown are
the measurements from \cite{ChepelLeff1999} (dark red), \cite{ArneodoLeff2000} (red),
\cite{AkimovLeff2002} (dark blue), \cite{AprileLeff2005} (purple), \cite{AprileLeff2009} (black),
\cite{PlanteLeff2011} (orange), and \cite{ManzurQy2010} (dark green). The two gray points from 
\cite{ChepelLeff1999} are not included in the fit. Absolute light yield $L_y$ is calculated by assuming 
a 63 photons/keV yield from $^{57}$Co at zero field, calculated in \cite{NESTpaper2}. Calculation of the error 
band is explained in the text. 
The maximum close to 100 keV is due to the Penning effect, which increases with energy according to Eq 6. 
While outside the region
of interest for WIMP searches and coherent neutrino scattering experiments, it is included for completeness 
and will be useful for comparison to future data. }
\label{fig:Leff}
\end{figure}

\begin{figure}[t]
\centering
\includegraphics[width=0.47\textwidth]{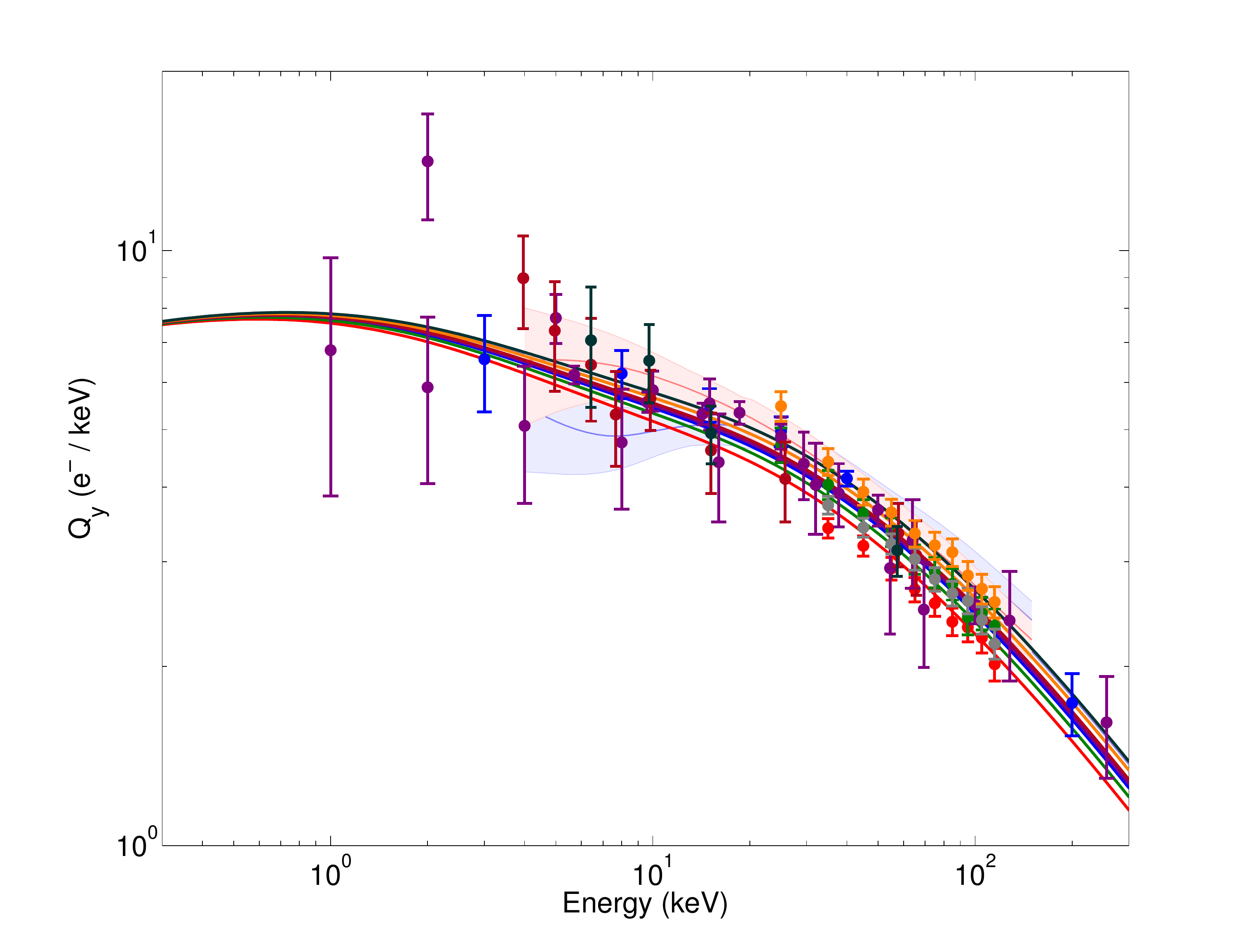}
\caption{(color online) All charge yield data included in the fit, plotted with best fit results from this work. Each color represents
a different applied electric field. Many detectors use applied electric fields to collect ionization electrons.  These data are at 100 V/cm (red) \cite{ColumbiaQy2006}, 270 V/cm (green) \cite{ColumbiaQy2006}, 
530 V/cm (blue) \cite{Xenon100Qy2013}, 730 V/cm (purple) \cite{SorensenQy2009, SorensenQy2010, XENON10Qy}, 
1000 V/cm (dark red) \cite{ManzurQy2010}, 2000 V/cm (orange) \cite{ColumbiaQy2006}, 2030 V/cm (gray)  
\cite{ColumbiaQy2006}, 3400 V/cm (blue band) \cite{HornQy2011}, 3900 V/cm (orange band) \cite{HornQy2011}, 
and 4000 V/cm (dark green) \cite{ManzurQy2010}.}
%\caption{Charge yield calculated at 730 V/cm with statistical error band. Also shown are data from
%\cite{SorensenQy2009} (red), \cite{SorensenQy2010} (green), and \cite{XENON10Qy} (orange). }
\label{fig:QyPlot}
\end{figure}

The best fit values and corresponding confidence intervals constructed from the MCMC are given in Table
\ref{tab:fits}, and the fits to the global data set are shown in Figures \ref{fig:DahlPlot}, \ref{fig:Leff}, and \ref{fig:QyPlot}. The mean values and confidence intervals are obtained
by marginalizing the MCMC sample set in each parameter and measuring the mean value and 34\% of the sample space on either
side of the mean. In fitting, the $\chi^2$/d.o.f. is calculated to allow comparison between models. We calculate (327 points - 9 free parameters - 1) = 317 degrees of freedom. The 
best-fit gives $\chi^2 / \text{d.o.f.} = 1.33$.

\begin{figure}[ht]
\centering
\includegraphics[width=0.498\textwidth]{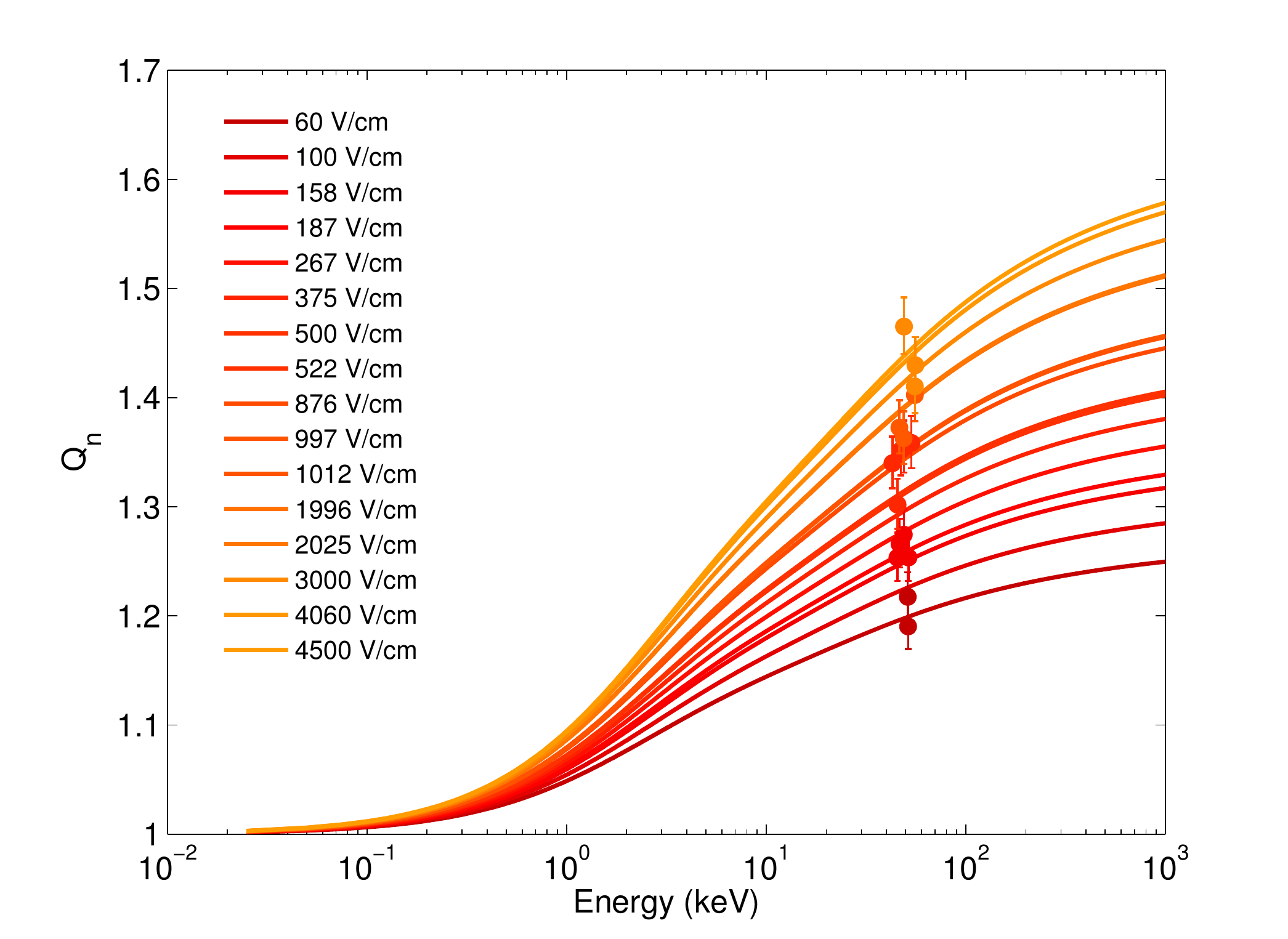}
\caption{(color online) Charge yield relative to that as applied electric field approaches 0, $Q_n = \mathcal{Q}_y ( F) / \mathcal{Q}_y (F_0)$. This quantity is calculated using the best-fit model. The measurements shown are not included in the fit and are only plotted here to compare with the model.
Existing measurements are at 56.5 keV, and are artificially offset for clarity.
Even with no drift field, some ionization electrons escape the interaction site \cite{NESTpaper}, but the enhancement of the ionization signal with higher applied electric field is shown explicitly here.
Data from \cite{ColumbiaQy2006}. To improve agreement, we use $F_0 = 7~\text{V/cm}$, well within the best-fit confidence interval of this parameter.}
\label{fig:Qn}
\end{figure}

 \begin{figure}[ht]
\centering
\includegraphics[width=0.47\textwidth]{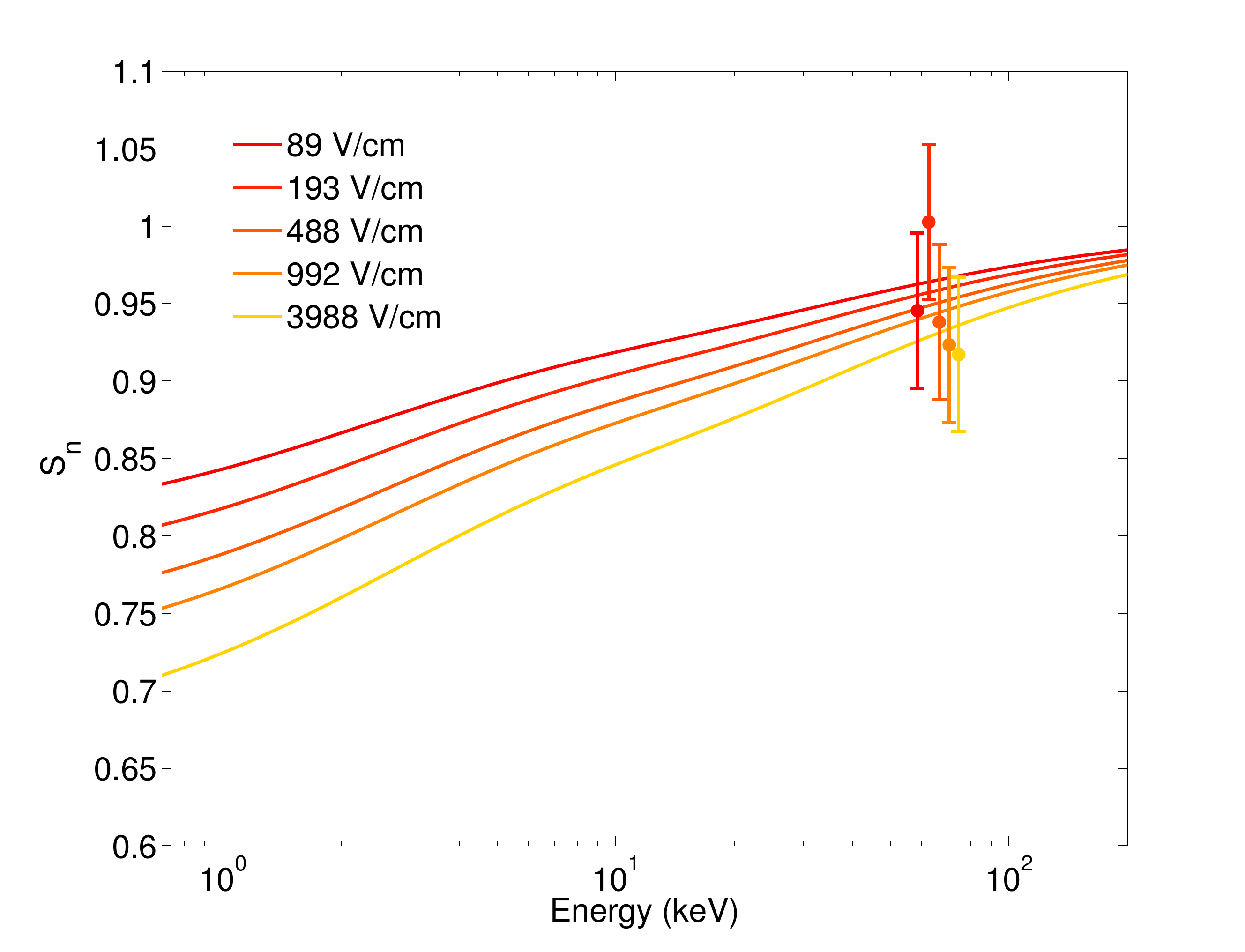}
\caption{(color online) Relative scintillation yield $S_n$ at different values of external electric field. This quantity is calculated from the best-fit model by 
dividing $n_{ph}$ at the given field by $n_{ph} (F_0)$. Measurements from \cite{ColumbiaQy2006} are not included in the global fit, but are plotted here to compare with the model. All measurements are at 56.5 keV, but are artificially offset for clarity. }
\label{fig:Sn}
\end{figure}

We note that the most tightly constrained parameter is the Lindhard $k$ factor. As described above, earlier estimates
place this value at $k = 0.166$ or $k = 0.110$ \cite{SorensenDahl}. The best fit obtained here lies in between
these two values, and is constrained to within 3\%.

The scaling coefficient $\alpha$ of $N_{ex}/N_i$ is of O$\left(1\right)$. This is in contrast with electronic recoils, in which $N_{ex}/N_i$ is found to be $< 0.2$ \cite{QingLin, AprileAntiCorr, DokeNex}. The scaling coefficient $\gamma$ of  $\varsigma$ is O$\left(0.01\right)$. Our fits for both coefficients agree with work in \cite{DahlThesis} and \cite{SorensenDahl}.  
%These values are $\alpha$ and $\gamma$ in our fit, respectively. 

The Penning quenching parameters, $\eta$ and $\lambda$, can also be compared to previous work. As stated above, the theoretical value for $\eta$ is 3.55, well within our confidence interval in our fit. The parameter $\lambda$ is expected to be $1/2$, significantly lower than our best fit. However, we are in agreement with the energy dependence of the stopping
power at high energies calculated by SRIM, shown in Fig 4.11 of \cite{DahlThesis}, which approaches $\sim1$ at high energies. Because this quenching has a
small effect except at high energies, we accept this result.

The effective zero-field nuisance parameter, $F_0$ is best-fit by 1.03 V/cm, with a 68\% confidence interval extending from arbitrarily close to
0 to 14.9 V/cm. As discussed in Sec. \ref{subsec:Parameterizing}, this is consistent with the constraints of experiments to date.

\subsection{\label{subsec:AbsoluteYields} Absolute Scintillation and Electron Yields}
One of the most important goals of this work is to produce a model that predicts absolute yields,
rather than remaining within the traditional paradigm of relative yields. Historically, measurements of the scintillation 
light production 
from nuclear recoils have been made relative to the scintillation light produced by the 122 keV $\gamma$-ray from $^{57}\text{Co}$ with no applied electric field. 
While relative quantities remain
critical to our understanding, the need to project the performance of future detectors with different efficiencies necessitates accurate predictions
of absolute fundamental quanta. With this goal in mind, the best-fit model is implemented in the NEST code and used to 
calculate yields in units of photons and electrons per unit energy. 

In these units, the absolute light yield of liquid xenon is defined as 
\begin{equation}
L_y = \frac{n_{ph}}{E_0}  = \mathcal{L}_{eff} \cdot \frac{S_n}{S_e} \cdot \frac{n_{ph}(^{57}\text{Co}) }{122 \text{ keV}}
\end{equation}
where $S_n$ and $S_e$ are the scintillation reduction factors due to applied fields for nuclear recoils and
electronic recoils, and $n_{ph}(^{57}\text{Co})$
is the yield from the 122 keV $\gamma$-ray with no applied electric field. To compare with measurements in the literature,
we calculate $L_y$ at $F = F_0$ ($S_n$ = $S_e$ = 1) and plot it against the measurements of $\mathcal{L}_{eff}$ used in the fit. 
In our calculations, we use $n_{\gamma}(^{57}\text{Co}) / 122 \text{ keV} = 63 \text{photons / keV}$, a value calculated using NEST in \cite{NESTpaper2} and confirmed experimentally in several experiments \cite{MIXPaper}
and the $\mathcal{L}_{eff}$ data are scaled to give absolute yields. 
The resulting $L_y$ as a function of incident energy is shown in Fig. \ref{fig:Leff} with
all measurements used in this work.  Charge yield is calculated simply as
\begin{equation}
\mathcal{Q}_y = \frac{n_e}{E_0} 
\end{equation}
Figure \ref{fig:QyPlot} shows the NEST calculations vs. energy with all measurements
used in the fit. The field dependence of this quantity is emphasized in Fig. \ref{fig:Qn}. Finally, taking the calculated ratio 
$n_e / n_{ph} $ allows direct comparison to the measurements in \cite{DahlThesis}. This is shown in 
Fig. \ref{fig:DahlPlot}. 

The uncertainty band in Fig. \ref{fig:Leff} shows the statistical uncertainty
calculated using the MCMC sample set. Each of the $3\times10^6$ points contains a set $\vec{\theta}$ of 
the nine free parameters, corresponding to some unique incarnation of our model. At each energy,
the value of interest ($L_y$ or $\mathcal{Q}_y$) is calculated for all points and placed into
a histogram. The statistical uncertainty is then taken as the standard deviation of this histogram. This technique
naturally incorporates all correlations between parameters. The width
is constrained simultaneously by all light and charge data in the fit, and is further constrained by the choice
of model, resulting in small uncertainties compared to the apparent spread in the data. Uncertainties on charge yield are
comparable, but are not shown for the sake of clarity.

The relative effects of applied electric fields on scintillation and ionization yields are shown in Figures \ref{fig:Qn} and \ref{fig:Sn}. Charge yield increases and scintillation decreases with higher applied fields, showing the suppression of recombination of electrons and ions. Our model is compared to available data that has not been included in the global fit.

The best-fit model can be modified within the framework presented as new data becomes available, such as the in-situ calibrations from the LUX experiment \cite{Verbus}.

\begin{figure}[!t]
\centering
\includegraphics[width=0.49\textwidth]{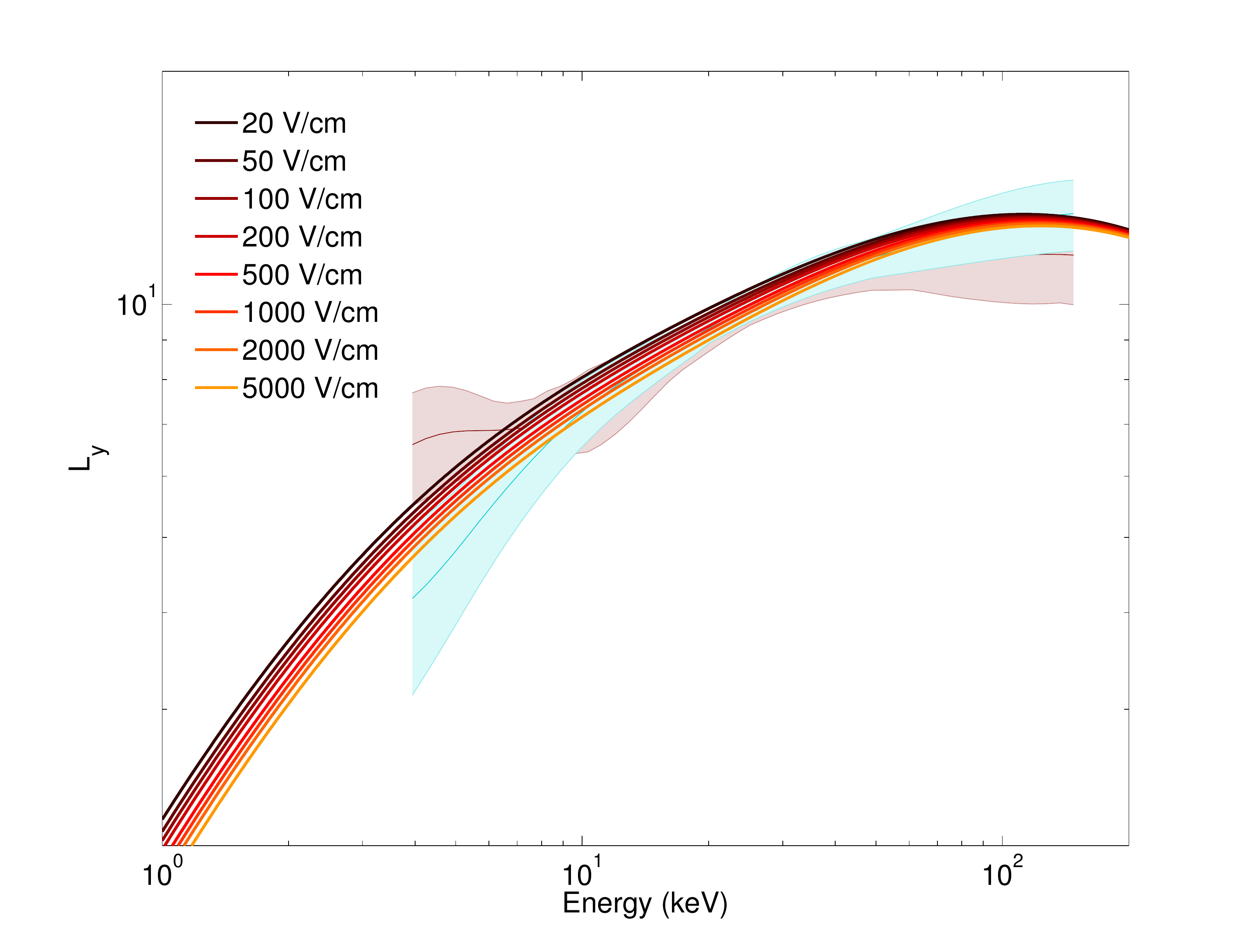}
\caption{(color online) Absolute scintillation yield $L_y$ at different values of external electric field. Also shown are the ZEPLIN-III indirect measurements
 from the first science run (red band) at 3900 V/cm and the second science run (blue band) at 3400 V/cm, found in \cite{HornQy2011}. This figure explicitly shows the predicted dependence of light yield on the applied electric field.  Stronger electric fields suppress recombination, and therefore reduce the scintillation signal at a given energy. }
\label{fig:LyAtFields}
\end{figure}

\section{\label{sec:Fluctuations}Adding Fluctuations To the Model}
It has been widely observed that the measured widths of  yields in liquid xenon are broader than the expectation from Poisson statistics \cite{XENON10Fluct, Bolotnikov}.  This is true for both light and charge yields, and in both the ER and NR samples.  It has further been shown
in \cite{DahlThesis} by subtracting detector effects that the variance in the yields is caused primarily by fluctuations in the
recombination process.  Consequently, combining the charge and light signals to solve for $E_0$ in Eq. 1 results in a combined energy scale that has a superior resolution than that of either signal alone \cite{SorensenDahl}.  In earlier versions of NEST, the broadening of fluctuations in electron-ion recombination was implemented as a complicated function of energy and field that had no physical justification.

Dobi has provided an improved treatment of this subject \cite{DobiThesis}, and his approach has been implemented in NEST to model
these fluctuations. Dobi observes a quadratic relationship between the variance (in units of quanta) and the number of ions 
produced in that event. Using electronic recoil data, he finds the approximate relationship:
\begin{equation}
\sigma_r^2 = \frac{1}{185} \times N_i^2
\label{eq:DobiFluct}
\end{equation}

The fluctuations are modeled using a Poisson distribution, modified such that the variance can be tuned in software to fit the data.
Physically, a binomial distribution is more appropriate for the application, but it is impossible to scale the variance to accommodate the observed width of fluctuations. In addition, the computational expense of binomial sampling is an impediment to efficient
simulation. While a Gaussian distribution could also have been used, it was decided against due to the non-zero probability of calculating negative quanta and the need to produce an integer number of quanta.

From Poisson statistics, we expect the variance to be equal to the number of electrons $n_e$, which can be written
as $(1-r) N_i$. If we assume the proportionality in Eq. \ref{eq:DobiFluct}, dividing the observed variance by the expected variance gives $\mathcal{F}_r = \frac{1}{185 (1-r)} N_i$.  Averaging the recombination probability for nuclear recoils over the range 1-100 keV gives $r \approx 0.5$,  which allows us to estimate the proportionality above to be $\mathcal{F}_r \approx 0.01 N_i$. 

\begin{figure}[t]
\centering
\includegraphics[width=0.47\textwidth]{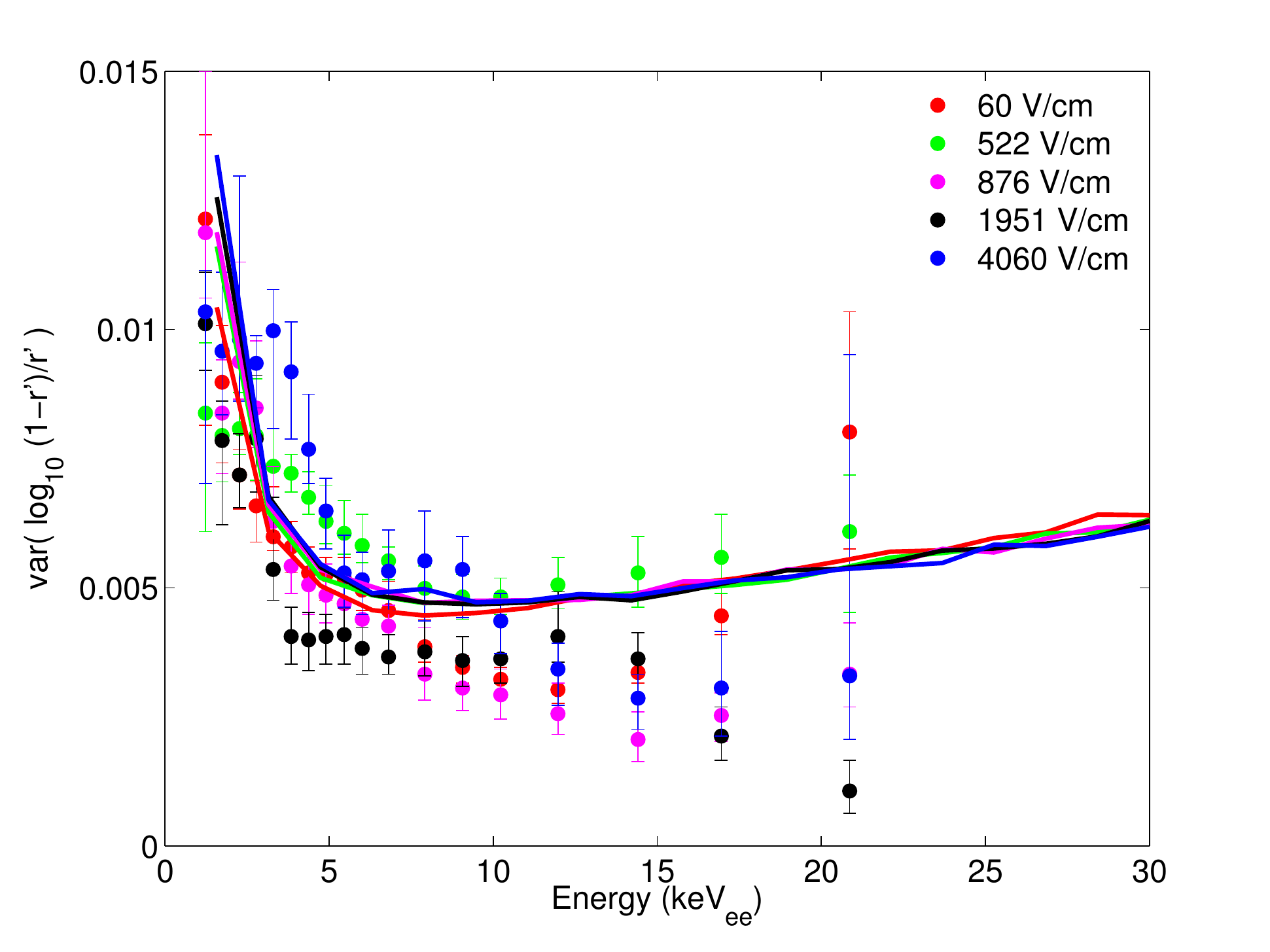}
\caption{(color online) The calculated variance in the ratio of electrons to photons, converted to a function of recombination fraction as done in \cite{DahlThesis}, plotted with the data contained therein.  These data are used to constrain the fluctuation model described in Sec. 
\ref{sec:Fluctuations}.  We observe very little field dependence in the prediction of the magnitude of fluctuations, and there is no conclusive field dependence in the measurement. Note, we use the variable $r'$, which is the recombination fraction as defined in \cite{DahlThesis}. This relies on the assumption of an exciton-to-ion ratio of
$N_{ex} / N_i$ = 0.06. The exciton-to-ion ratio in this work is different, so $r'$ is distinct but maps directly to the recombination fraction $r$ presented in this work. }
\label{fig:Fluct}
\end{figure}

To accommodate this model, we define a Fano-like factor for recombination fluctuations
\begin{equation}
\mathcal{F}_r = \mathcal{C} N_i
\end{equation}
to quantify the deviation of 
the observed fluctuations from the expected Poisson statistics.  To fit to nuclear recoil data, we treat $\mathcal{C}$ as a final, tenth free parameter.  A fit to data from \cite{DahlThesis} yields a value $\mathcal{C}=0.0056$, given in Table \ref{tab:fits}. The data used and the fit are shown in Fig. \ref{fig:Fluct}.

\begin{figure}[t]
\centering
\includegraphics[width=0.49\textwidth]{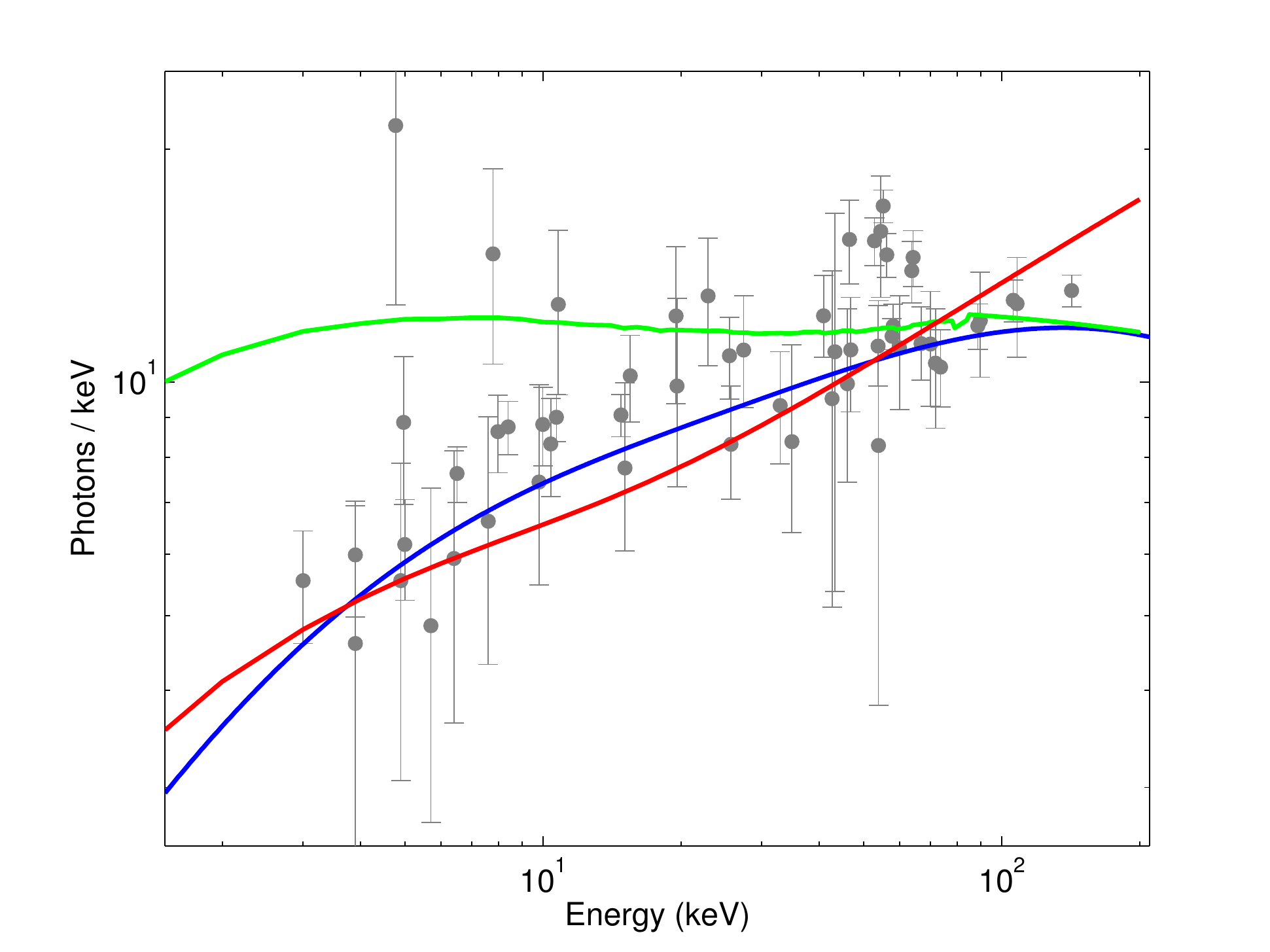}
\caption{(color online) All measurements of $\mathcal{L}_{eff}$ plotted with two alternative models of nuclear stopping power
from \cite{Bezrukov}. The solid lines are the best fit (blue), Ziegler et al. (red), and Lenz-Jensen (green). }
\label{fig:BezrukovLeff}
\end{figure}

\begin{figure}[th]
\centering
\includegraphics[width=0.49\textwidth]{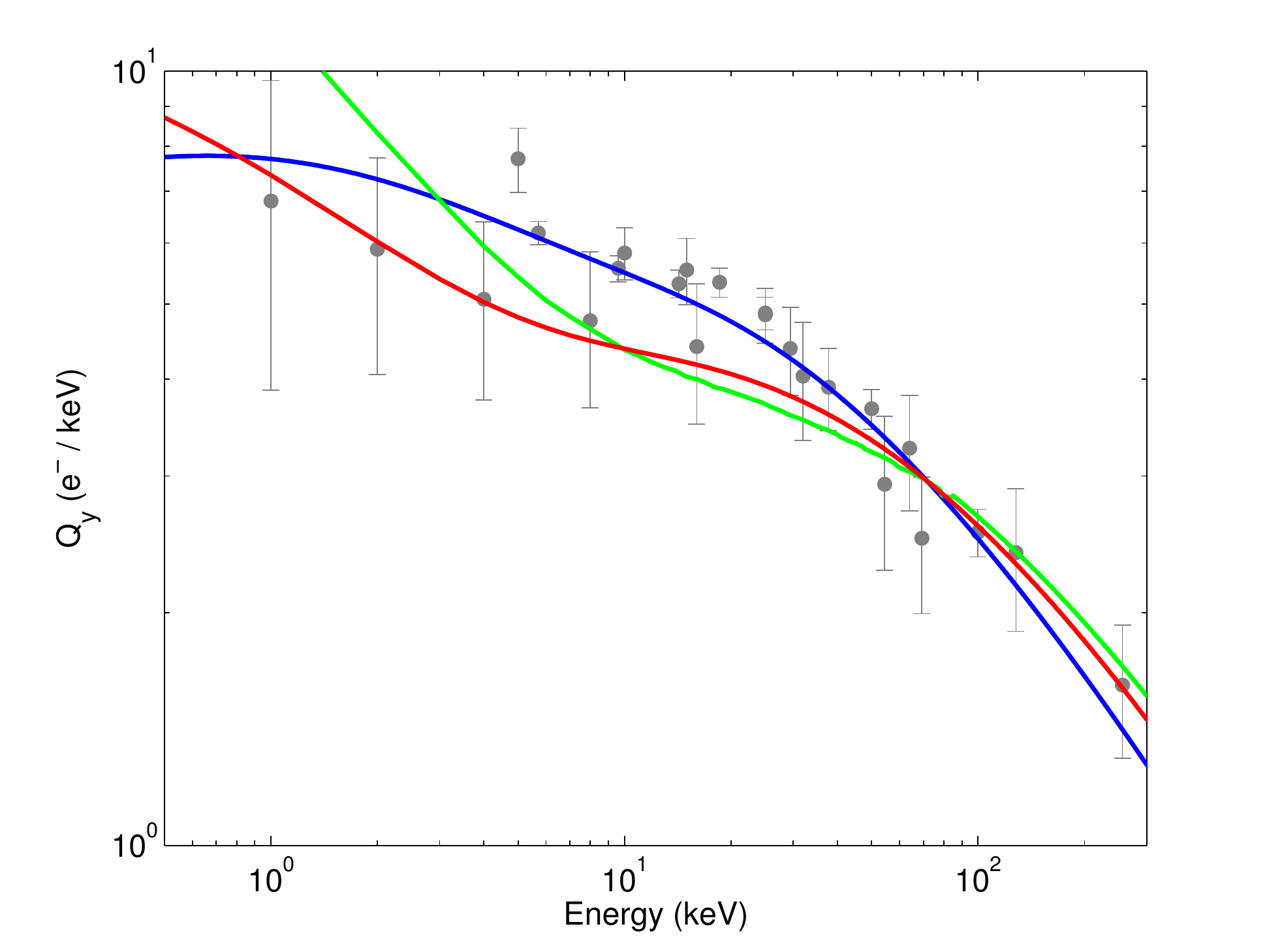}
\caption{(color online) Measurements of $\mathcal{Q}_y$ at a drift field of $730$ V/cm, plotted with alternative
models of nuclear stopping power from
\cite{Bezrukov}. This field was chosen as an example due to the existence of multiple analyses.
The solid lines are the best fit (blue), Ziegler et al. (red), and Lenz-Jensen (green). }
\label{fig:BezrukovQy}
\end{figure}

\section{\label{sec:Discussion}Discussion of Alternative Models}
The deposition of energy in liquid xenon is a complicated process, and there are many theories
that seek to calculate observable quantities such as charge and light yields from first principles.
Several of these are described in detail by Bezrukov et al. in \cite{Bezrukov}, in which the authors
compare several models for the distribution of energy into nuclear and electronic excitations. The models
presented can act as alternatives to Lindhard.  In addition, recent work by Mu et al. suggests a new alternative via an
 extrapolation of the electronic stopping power in gaseous xenon measured in \cite{Fukuda}. In this section, 
 we discuss how these models compare when using the global approach 
taken in this work, and show that the model described in Sec. \ref{subsec:Theory} is the best fit to existing data. 

Bezrukov et al. study the effect of different calculations of the nuclear stopping power $s_n$, which they
define to be proportional to the probability that a recoiling xenon atom will scatter elastically from another xenon atom.
This is used in conjunction with the electronic stopping power $s_e$  to determine the fraction of energy 
a recoiling xenon atom loses to observable electronic excitation in the detector,
\begin{equation} 
L = \frac{s_e}{s_n + s_e}
\end{equation}
The Thomas-Fermi model, used by Lindhard in the treatment adopted in Sec. \ref{subsec:Theory}, is compared to two
alternative models of $s_n$; that due to Ziegler et al. and that due to 
Lenz and Jensen. We incorporate both into our simulated annealing fitter, replacing the free parameter $k$ from
Lindhard's model with an overall scaling of the quantity $L$. This allows the resulting curve to shift
vertically to fit the world's data. These alternative models are used to produce 
the $L_y$ and $\mathcal{Q}_y$ curves in Figs. \ref{fig:BezrukovLeff} and \ref{fig:BezrukovQy}.  The best-fit 
$\chi^2/\text{d.o.f}$ values obtained using the simulated annealing
fitter were 2.88 and 4.32 respectively, compared to $1.33$ in the current work.

In addition, Bezrukov et al. offer two possible low-energy correction factors corresponding to enhancement
or suppression of the electronic stopping power. This affects the number of total
quanta generated at low energies. To test these, we remove the energy dependence on the exciton-to-ion ratio
and introduce these corrections into the Lindhard factor. These corrections are alternative ways of expressing
an energy-dependent yield, with the current energy dependence of $N_{ex} / N_i$ corresponding to a suppression of the 
light yield. We find a best-fit $\chi^2/\text{d.o.f.} 
= 2.43$ for the Bezrukov enhancement model, and $\chi^2/\text{d.o.f} = 1.84$ for suppression. 

The calculations due to Mu et al. predict a quenching factor $L$
that is significantly smaller than previous work \cite{MuJiQy, MuJiLeff}. In the global context, this results in fewer quanta 
produced for a given energy deposition. However, from Eq.~1 we see that this can be reconciled by assuming a smaller $W$. Mu et al. favor the value $W = 9.76~\text{eV}$, for which they 
 cite the experimental measurement in \cite{Seguinot}. We find reasonable agreement with the global fit in the energy regime where experimental data exist ($E_0$ $>$ 1 keV) by 
assuming $W = 8.3~\text{eV}$ (approximately $2\sigma$ below the measured $W$ in \cite{Seguinot}), and we can bring the Lindhard model
into agreement with this smaller $W$ by assuming $k = 0.07$. This energy is closer to the true ionization and scintillation energies of xenon, and may appear favorable from this perspective. However, the Lindhard model is found to have excellent agreement in germanium when using the average energy required to produce quanta rather than the true band gap energy \cite{Barbeau}, so in the implementation of this model, $W = 13.7\, \text{eV}$.

\section{\label{sec:Summary}Summary}
We have presented here a model of light and charge yields from nuclear recoils in liquid xenon, 
constrained simultaneously using measurements of both quantities. 
This approach incorporates an anti-correlation between the two and helps break degeneracies
between quantities that can independently affect one or the other. 
We are able to obtain a better constrained mean for the semi-empirical NEST model, and find that it provides
a better fit than alternatives suggested in the literature. 

Historically, experimental results are reported using assumptions based on individual calibrations. The differences in these calibration measurements can complicate the interpretation of physics results and comparison to past and present experiments.  The global analysis technique that we develop here is a step towards distilling all of these calibration measurements into a comprehensive and consistent picture of liquid xenon response to radiation.  Such an understanding will facilitate the interpretation of experimental results in a broader context.

Looking forward, the extensive treatment of the energy- and field-dependence of yields contained in this work can guide the analyses and designs of existing and future experiments. For example, our model shows that ionization remains substantial at very low energies, implying that the charge channel can continue to be used effectively as experimental thresholds are lowered. In terms of experimental design, the model can be used, for instance, to benchmark requirements for applied electric fields in detectors. The NEST package is currently used in simulations of the next-generation dark matter experiment LUX-ZEPLIN \cite{LZ_CDR} to model response as the detector is designed.  Our results from the global analysis improve the accuracy and confidence in the underlying model, and can assist the optimization of experiments to maximize sensitivity and background rejection.

\section{\label{sec:Acknowledgements}Ackowledgements}
The authors thank the LUX Collaboration, for ideas and discussions on the present work.
We particularly thank Attila Dobi, Carmen Carmona, James Verbus, Kevin O'Sullivan, Dan McKinsey, Markus Horn, Evan Pease,
and Rick Gaitskell for comments and suggestions on the model. We thank Richard Ott for helpful discussions of Markov Chain Monte Carlo techniques. Brian Lenardo is supported by the Lawrence Scholars Program at the 
Lawrence Livermore National Laboratory. Lawrence Livermore National Laboratory is operated by Lawrence Livermore 
National Security, LLC, for the U.S. Department of Energy, National Nuclear Security Administration under Contract DE-AC52-07NA27344. 
This work was supported by U.S. Department of Energy grant DE-FG02-91ER40674 at the University of California, Davis, as well 
as supported by DOE grant DE-NA0000979, which funds the seven universities involved in the Nuclear Science and Security 
Consortium. LLNL-JRNL-664499.

% Can use something like this to put references on a page
% by themselves when using endfloat and the captionsoff option.
\ifCLASSOPTIONcaptionsoff
  \newpage
\fi

% trigger a \newpage just before the given reference
% number - used to balance the columns on the last page
% adjust value as needed - may need to be readjusted if
% the document is modified later
%\IEEEtriggeratref{8}
% The "triggered" command can be changed if desired:
%\IEEEtriggercmd{\enlargethispage{-5in}}

% references section
% can use a bibliography generated by BibTeX as a .bbl file
% BibTeX documentation can be easily obtained at:
% http://www.ctan.org/tex-archive/biblio/bibtex/contrib/doc/
% The IEEEtran BibTeX style support page is at:
% http://www.michaelshell.org/tex/ieeetran/bibtex/
%\bibliographystyle{IEEEtran}
% argument is your BibTeX string definitions and bibliography database(s)
%\bibliography{IEEEabrv,../bib/paper}
%
% <OR> manually copy in the resultant .bbl file
% set second argument of \begin to the number of references
% (used to reserve space for the reference number labels box)
\bibliographystyle{./IEEEtran}
\bibliography{./IEEEabrv,./NESTpaper}
\end{document}